\documentclass[10pt]{ismd08}
\usepackage{graphicx}
\usepackage{cite,./mcite}

\begin{document}
\title{ISMD08 \\ A Brief Introduction to the Color Glass Condensate and the Glasma}
\author{Larry McLerran}
\institute{RIKEN-BNL Center and Physics Department, Brookhaven National Lab., Upton, NY, 11973 USA\\  }
\maketitle
\begin{abstract}
I provide a brief introduction to the theoretical ideas and phenomenological motivation for the
Color Glass  Condensate and the Glasma.

\end{abstract}

\section{Introduction}

The purpose of this talk is to motivate the ideas behind the Color Glass Condensate and the Glasma.  As space is limited in such conference proceedings, the references below are not comprehensive, and the reader interested in a fuller documentation and an expanded discussion of the topics below are referred to a few  reviews where the original ideas are motivated, and where there are detailed references to the original literature.\cite{reviews}

\begin{figure}[htbp]
\begin{center}
\includegraphics[width=.99\textwidth] {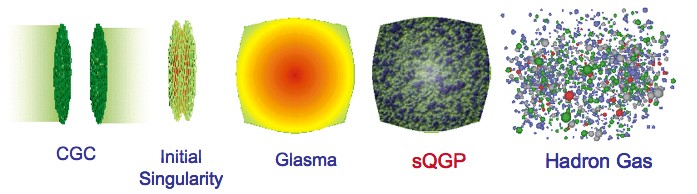} 
\end{center}
\caption{An artistic conception of high energy hadronic collisions}
\label{bass}
\end{figure}

The concepts associated with the Color Glass Condensate and the Glasma were generated to address at least three fundamental questions in particle and nuclear physics:
\begin{itemize}

\item{\it What is the high energy limit of QCD?}

\item{\it What are the possible forms of high energy density matter?}

\item{\it How do quarks and gluons originate in strongly interacting particles?}

\end{itemize}
The Color Glass Condensate is the high energy density largely gluonic matter that is associated with
wavefunction of a high energy hadron.\cite{mv}-\cite{jimwlkmueller}   It is the initial state in high energy hadronic collisions,
and its components generate the distributions of quarks and gluons measured in high energy
deep electron scattering from nuclei.  Almost instantaneously after a hadron-hadron collision, the nature of the gluonic matter changes its structure, and the Color Glass Condensate fields are transformed into longitudinal color electric and magnetic fields, the Glasma.\cite{weigert}-\cite{lappi}  These early stages are shown in Fig. \ref {bass}.
Later, the Glasma decays and presumably thermalizes and forms a Quark Gluon Plasma, which eventually itself decays into hadrons.\cite{theorywhitepaper}-\cite{whitepaper}  

\begin{figure}[htbp]
\begin{center}
\includegraphics[width=.80\textwidth] {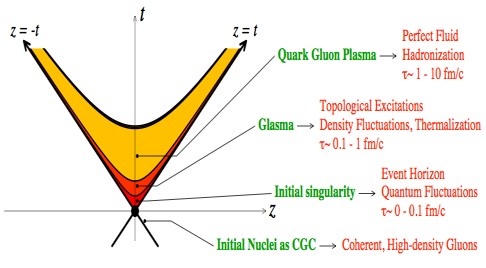} 
\end{center}
\caption{A space-time diagram for hadronic collisions}
\label{spacetime}
\end{figure}

A more detailed picture of the evolution of the matter produced in heavy ion collisions is shown in the space-time diagram of Fig. \ref{spacetime}  This picture demonstrates the close correspondence between the physics of hadronic collisions and that of cosomology.
There is an initial singularity and singularity along the light cone.  In the expansion, topological
excitations are generated, which I  will describe later.  These excitations have a correspondence with
the topological charge changing processes of electroweak theory, which may be responsible for
the baryon asymmetry of the universe.\cite{manton}-\cite{arnoldmclerran}  There is a 1+ 1 dimensional analog of Hubble expansion in hadronic collision, which corresponds to that of  3+1 dimensional Hubble flow in cosmology.\cite{shuryakanishetty}-\cite{bjorken}

I have purposely not tried to discriminate here between heavy ion collisions and hadronic collisions.
Of course the energy density,  and the validity of various approximations may depend upon the energy and the nature of particles colliding.  The physics should be controlled by the typical density
of produced particles, and when the size of the system $R$ becomes large compared to the typical inter-particle separation $d \sim \rho^{-1/3}$, one is justified in taking the large size limit.
This should ultimately happen at very high energies since the density of gluons rises as beam energy
increases, but it may be greatly enhanced by the use of nuclei.

The reason for this high density of gluons is that the size of a hadron, for example a proton, grows very slowly as a function of collision energy.  On the other hand, the number of gluons grows rapidly.
The lowest fractional momentum $x$ values probed at some collision energy are typically $x \sim \Lambda_{QCD}/E$,
so that the number of gluons at small x is a measure of the number of gluons appropriate for the description of a hadron at energy E.  
\begin{figure}[htbp]
\begin{center}
\includegraphics[width=.60\textwidth] {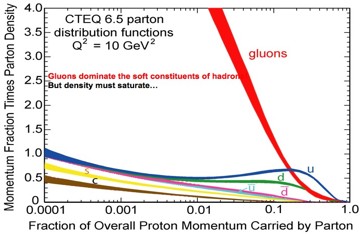} 
\end{center}
\caption{A distribution of quarks and gluons in  a hadron as a function of x}
\label{hera}
\end{figure}
Note that gluons dominate the wavefunction of a hadron for $x \le 10^{-1}$, shown in Fig. \ref{hera}.\cite{hera}-\cite{stasto}

\begin{figure}[htbp]
\begin{center}
\includegraphics[width=.40\textwidth] {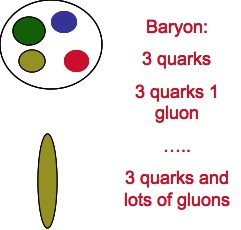} 
\end{center}
\caption{The Fock space states corresponding to a high energy hadron wavefunction.}
\label{fock}
\end{figure}

These gluons are associated with states in the high energy hadron wavefunction.  As shown in Fig. \ref{fock}, the Fock space components of a nucleon wavefunction have states with 3 quarks, and
3 quarks with arbitrary numbers of gluons and quark-antiquark pairs.  The part of the wavefunction with
3 quarks and a few quark-antiquark pairs dominates properties of the nucleon measured in intermediate energy processes and high energy processes at large $x$.  The part with many gluons controls typical
high energy processes.  These states have a very high density of gluons.

Before proceeding further, I want to review some of the kinematic variables useful for the high energy limit.  I introduce light cone variables associated with time, energy, longitudinal coordinate and momentum,
\begin{eqnarray}
  x^\pm  & = & (t  \pm z)/\sqrt{2} \nonumber \\
  p^\pm  & = & (E \pm p_Z)/\sqrt{2}
\end{eqnarray}
The  dot product is $x\cdot p = x_T \cdot p_T - x^+p^- - x^- p^+$.  The uncertainty principle is
$x^\pm p^\mp \ge 1$. A longitudinal boost invariant proper time is $\tau = \sqrt{t^2 - z^2}$ and corresponding transverse mass is $m_T = \sqrt{p_T^2 + M^2}$.  There are several types of rapidity corresponding to different choices of space-time or momentum space variable.  Using definitions and the uncertainty principle, we see that up to uncertianties of order one unit of rapidity, all of these rapidity variables are the same,
\begin{equation}
 y = {1 \over 2 } ln (p^+/p^-) = ln(p^+/m_T) \sim -ln(x^-/\tau) = -{1 \over 2} ln(x^+/x^-) = -\eta
\end{equation}

\section{The Color Glass Condensate}

The Color Glass Condensate is the matter associated with the high density of gluons
appropriate for the description of the wavefunction of a high energy hadron.  In the following,
I motivate the CGC, and discuss phenomenological implications.

\subsection{The Color Glass Condensate and Saturation}

As gluons are added to a high energy particle wavefunction, where do they go?  The size of a hadron is roughly constant as energy increases.  If we add gluons  of fixed size then surely at some energy scale
these gluons will closely pack the area of a hadron.  Repulsive interactions of order $\alpha_s$
will become important and the packing will shutoff when the density is of order $1/\alpha_s$.  How can more gluons be packed into the hadron?  Resorting to an analogy with hard spheres, we can pack in more gluons if their size is small.  They can fit into the holes between the closely packed gluons of larger size.
This process can go on forever, packing in gluons of smaller and smaller size  as the energy increases.
There is a characteristic momentum scale $Q_{sat}$ which corresponds to the inverse size scale of
smallest gluons which are closely packed.  The saturation momentum, $Q_{sat}$ grows as the energy increases.  Note that saturation does not mean the number of gluons stopped growing, only that for gluons of size larger than $1/Q_{sat}$, they have stopped growing. 
\begin{figure}[htbp]
\begin{center}
\includegraphics[width=.40\textwidth] {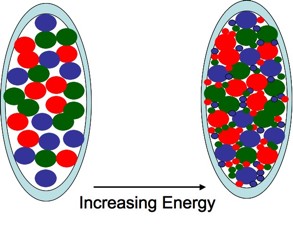} 
\end{center}
\caption{Gluons being added to the wavefunction of a hadron as energy increases.}
\label{saturation}
\end{figure}

We can now understand the name Color Glass Condensate.  The word color comes from the color of the gluons.  The word condensate comes from the high density of gluons.  The phase space density of gluons is
\begin{equation}
  {{dN} \over {dyd^2p_Td^2x_T}} = \rho
\end{equation}
There is some effective potential which describes the gluons.  At low density, $V \sim - \rho$, since the system wants to increase its density.  On the other hand, repulsive interactions balance the inclination to condense, $V_{int} \sim \alpha_s \rho^2$.  These contributions balance one another when
$\rho \sim 1/\alpha_s$.  The phase space density measures the quantum mechanical density of states.
When $Q_{sat} >> \Lambda_{QCD}$, the coupling is weak, and the phase space density is large.  The gluons are in  a highly coherent configuration.  The density scaling as the inverse interaction 
strength $1/\alpha_s$ is characteristic of a number of condensation phenomena such as the Higgs condensate, or superconductivity.

The word glass arises because the gluons evolve on time scales long compared to the natural time scale $1/Q_{sat}$.   The small $x$ gluons are classical fields produced by gluons at larger values of $x$.  These fast gluons have their time scale of evolution time dilated relative to their natural one.  This scale of time evolution is transferred to the low $x$ gluons.  This means that the low $x$ gluons can be approximated by static classical fields, and that different configurations of gluons which contribute to
the hadron wavefunction can be treated as a non-intrefering ensemble of fields.  These are properties of spin glasses.
\begin{figure}[htbp]
\begin{center}
\includegraphics[width=.20\textwidth] {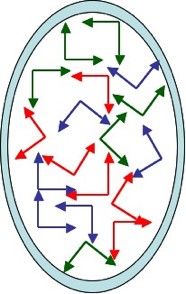} 
\end{center}
\caption{The color electric and color magnetic fields associated with the Color Glass Condensate.}
\label{fields}
\end{figure}

The configuration of color electric and magnetic fields of the CGC are determined by kinematics.  Let
the hadron be a thin sheet located a $x^- \sim 0$.  The fields will be slowly varying in $x^+$.  Therefore
$F^{i-}$ is small, $F^{i+}$ is big and the transverse field strengths $F^{ij}$ are of order one.  This means
that $\vec{E} \perp \vec{B} \perp \hat{z}$, so that the fields have the form of Lorentz boosted Coulomb fields of electrodynamics.  They are shown in Fig. \ref{fields}.  They have random color, polarization
and density.  The theory of the CGC determines the spectrum of these fluctuations.

\subsection{The Renormalization Group}

The spectrum of fluctuations is determined by renormalization group arguments.  The separation between what we call low $x$ and large $x$ gluons is entirely arbitrary.  The way we treat gluons below some separation scale is as classical fields with small fluctuation in the background.  For gluons with $x$ larger than the separation scale, we treat them as sources.  If we shift the separation scale to a lower value, we need to integrate out the fluctuations at intermediate scales, and they become the sources
at high $x$.  This integration is necessary since the fluctuations generate corrections to the classical theory of the form $ \alpha_s ln(x_0/x)$ where $x_0$ is the separation scale and $x$ is typical of the gluons.  If $x$ is too small compared to $x_0$, then the classical filed treatment does not work.  One has
to integrate out fluctuations recursively by the method of the renormalization group to generate a theory
on the scale of interest.  In this way, the fluctuations become new sources for fields at yet lower values of $x$.

This method of effective field theory was developed to treat the CGC.  It was found that the evolution
equations are diffusive, and have universal solutions at small x.  No matter what hadron one starts with, the matter one eventually evolves to is universal!.  The diffusive nature of the evolution means that the number of gluons and the saturation momentum itself never stop growing.  At high enough energy,
the coupling therefore becomes weak, although because the field strengths are large, the system is non-perturbative.

The renormalization group predicts the dependence of the saturation momentum on energy.\cite{kazularry}-\cite{dion}  In lowest order, it predicts the power law dependence on $x$ but with too strong a dependence.  This is corrected in higher order, and generates a reasonable description about what is known from experiment.\cite{reviews}

The renormalization group analysis also describes limiting fragmentation and small deviations from in it,
in accord with experimental observation.\cite{phobos}-\cite{gelisstasto}

\subsection{The CGC Provides and Infrared Cutoff}

The CGC acts as in infrared cutoff when computing the total multiplicity.\cite{venugopalan}, \cite{hirano}   For momentum scales $p_T \ge Q_{sat}$ a produced particle sees individual incoherent partons, and the results of ordinary perturbation theory which uses incoherent parton distributions should apply.  If the rapidity distribution of gluons in a hadron was roughly constant, then the distribution of gluons would follow be $dN/d^2p_Tdy \sim 1/p_T^2$ . At small $p_T \le Q_{sat}$, the gluon distribution is cutoff since a produced parton sees a  coherent field produced by a distribution of sources which is color neutral on the scale $1/Q_{sat}$.
This reduces the strength $dN/d^2p_Tdy \sim constant/\alpha_s$, up to logarithms.  We therefore obtain
\begin{equation}
  {{dN} \over {dy}} \sim {1 \over \alpha_s} \pi R^2 Q_{sat}^2
\end{equation}
In hadron-hadron collisions, the $1/p_T^4$ spectrum is also cutoff at $p_T \sim Q_{sat}$, so it too has the form above.

\subsection{The Total Cross Section}

Assume that the distribution in impact parameter and rapidity factorizes
\begin{equation}
 {{dN} \over {dyd^2r_T}} = Q_{sat}^2 e^{-2m_\pi r_T}
\end{equation}
The dependence above in the impact parameter profile is correct at large $r_T$ since the cross section
should be controlled by isospin zero exchange.  The total cross section measured by some probe
is determined by the maximum radius for which $Q_{sat}^2 e^{-2m_\pi b_T} \sim constant$.  Using the phenomenological parameterization of the saturation momentum $Q_{sat}^2 \sim e^{\kappa y}$, we see that $b_T \sim y$, so that the total cross section behaves as
\begin{equation}
  \sigma \sim b_T^2 \sim y^2 \sim ln^2(E/\Lambda_{QCD})
\end{equation}
The cross section saturates the Froissart bound.\cite{froissart}-\cite{ferreiro}  (This becomes modified somewhat when the gluon distribution is computed using a running coupling constant.)

\subsection{Qualitative Features of Electron-Hadron Scattering}
\begin{figure}[htbp]
\begin{center}
\includegraphics[width=.50\textwidth] {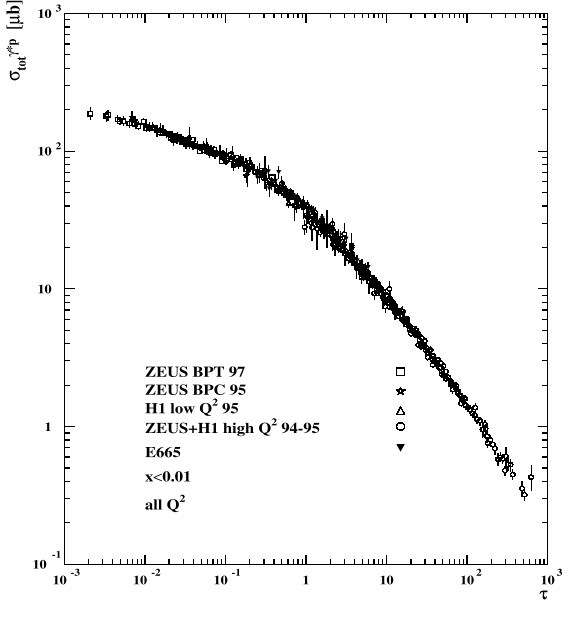} 
\end{center}
\caption{Geometric Scaling of  $\sigma_{\gamma^* p}$}.
\label{geometric}
\end{figure}

The cross section $\sigma^{\gamma^* p}$ is measured in deep inelastic scattering.  In the CGC,
this cross section is determined by computing the expectation values of electromagnetic currents in the background of the CGC fields.  On dimensional grounds, it is only a function of $F(Q^2/Q_{sat}^2)$.  In the theory of the CGC, the only energy dependence appears through the dependence of the saturation momentum upon energy.  There is no separate energy dependence.  This means that the data on deep inelastic  scattering should scale in $\tau = Q^2/Q_{sat}^2$  The data on $\sigma^{\gamma^* p}$ for $x \le 10^{-2}$ is shown in Fig. \ref{geometric}.  It clearly demonstrates this scaling.  The data for $x \ge 10^{-2}$ does not demonstrate the scaling.

It is a little strange that the data has such scaling for $\tau >> 1$.  This region is far from that of saturation.
It can nevertheless be shown that for $Q^2 << Q_{sat}^4/\Lambda^2_{QCD}$ that such scaling occurs,
although this is the region where a DGLAP analysis is valid.  The saturation momentum in this region appears as a result of a boundary condition of the DGLAP evolution.  It is also possible to compute the
structure functions for deep inelastic scattering in a way which includes both the effects of the CGC and
properly includes DGLAP or BFKL evolution, and good descriptions of the data are obtained.  The weakness of this analysis is of course that at Hera energies, the values of $Q_{sat}^2$ are not so large.

Diffractive deep inelastic scattering may also be analyzed using techniques of saturation and the CGC.  One can obtain a good description of diffractive structure functions.  The CGC and the impact parameter profile of hadrons are inputs for such computations.\cite{stasto}, \cite{gb}-\cite{kowalski}

\subsection{The CGC and Shadowing}
\begin{figure}[htbp]
\begin{center}
\includegraphics[width=.55\textwidth] {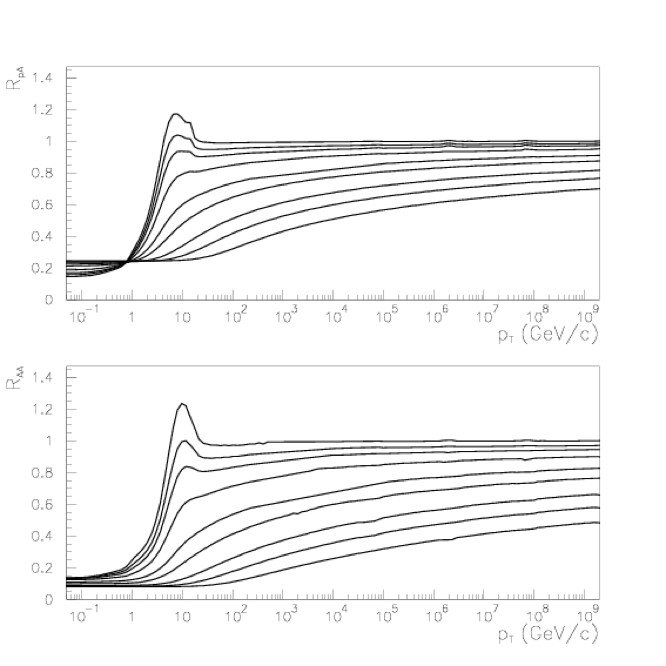} 
\end{center}
\caption{The ratios $R_{pA}$ and $R_{AA}$ of particles produced in $pA$ and $AA$ collisions to that
in $pp$ as a function of the transverse momentum of the produced particle.  The curve at the largest value of $x$ is the top curve, and evolution to smaller values of $x$ proceeds by moving further down. }
\label{cronin}
\end{figure}

The CGC provides a theory of shadowing for hadron nucleus interactions. \cite{kovner3}-\cite{jamal} Such shadowing can
be measured in lepton-nucleus scattering, $pA$ scattering or $AA$ scattering.  There are two competing effects.  The first is that the effects of a gluon propagating in a background field are distorted by the background field.  Such a field is stronger in a nucleus than in a proton.  These fields generate more momentum in the gluon distribution function at intermediate momentum at the expense of gluons at low momentum.  This will cause a Cronin peak in the ratios $R_{AA}$ and $R_{pA}$ at intermediate 
$p_T$.  It can be shown that this effect reflects multiple scattering of a projectile interacting with a target.
Such a peak is shown in Fig. \ref{cronin}.  Evolution of the distribution function to low values of $x$ is also affected by the CGC.  The CGC saturation momentum acts as a cutoff on the evolution equations.  Since the saturation momentum is larger for a nucleus than for a proton, as the distribution functions are evolved to smaller values of $x$, they are overall suppressed in a nucleus relative to a proton.  This is also shown in Fig. \ref{cronin}, where the distribution functions at different values
of $x$ are shown as a result of a computation for the CGC.

At RHIC, distributions of particles were measured in $dA$ collisions.\cite{brahms}  In the forward region of such collisions, values of $x \sim 10^{-3}$  of the gold nucleus were measured.  At large values of $x$, a clear Cronin enhancement was found for intermediate $p_T$.  At small $x$, the distributions were suppressed,
as predicted by the CGC.  Also at small $x$, as one increased the associated multiplicity in the collision
corresponding to more central collisions, there was less suppression at large $x$ and more suppression at small $x$.  One could describe these observations within the CGC framework.  The conclusions from such an analysis are somewhat limited by the kinematic limitations of the energy involved. 

\section{The Glasma}

\begin{figure}[htbp]
\begin{center}
\includegraphics[width=.55\textwidth] {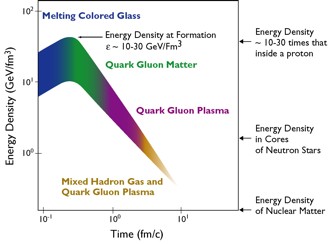} 
\end{center}
\caption{The various stages of heavy ion collisions.}
\label{ludlam}
\end{figure}

In Fig. \ref{ludlam}, I present the various stages of heavy ion collisions.  
The times and energy densities on this plot are appropriate for gold-gold collisions at RHIC.
The times are measured after the collision is initiated.  There is at late times a Quark Gluon Plasma, which is the matter after it has had time to thermalize.  Quark Gluon Matter exists at  intermediate times when the degrees of freedom are not highly
coherent, nor thermalized, and can be thought about as quarks and gluons.  The earliest times are when the fields are highly coherent, and most of the energy is in coherent field degrees of freedom,
not in the degrees of freedom of incoherent quarks and gluons.  The matter at earliest times is called
the Glasma, since it is neither a Quark Gluon Plasma, nor a Color Glass Condensate, but has features of both.  As we shall see below, almost instantaneously after the collision, the field configurations of the CGC, which are transverse, change into those of the Glasma, which are longitudinal.  It is the creation and decay of these Glasma fields which will be the subject of the next subsections.

\section{Colliding Sheets of Colored Glass}
\begin{figure}[htbp]
\begin{center}
\includegraphics[width=.40\textwidth] {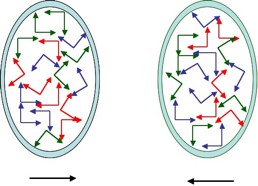} 
\end{center}
\caption{Colliding sheets of colored glass..}
\label{collidingsheets}
\end{figure}

Two hadrons in collision are visualized sheets of colored glass.\cite{weigert}-\cite{lappi}  They solve classical Yang-Mills equations before and after the collision.  These  classical fields change their properties dramatically
during the collision.  Prior to the collision, they are transverse to the direction of motion and confined to the region of thin sheets.  In the time it takes light to travel across the thin sheets, the sheets become charged with color electric and color magnetic charge.  This results in longitudinal color electric
and color magnetic fields between the sheets.  The typical strength of these fields is of order 
$Q_{sat}^2/\alpha_S$, and the typical variation in the transverse direction is  on a size scale $1/Q_{sat}$ These sheets have a large topological charge density since $\vec{E} \cdot \vec{B} \ne 0$.  It can be shown that such a description follows from first principles within the formalism of the Color Glass Condensate, and satisfies factorization theorems, similar to those in the parton model.\cite{factorization}

\begin{figure}[htbp]
\begin{center}
\includegraphics[width=.40\textwidth] {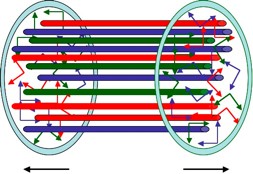} 
\end{center}
\caption{The Glasma as it appears in early stages of hadronic collisions.}
\label{glasma}
\end{figure}

The initial distribution of longitudinal color electric and magnetic fields and their evolution until thermalization is referred to as the Glasma.  It has properties such as the coherent strong fields that
are similar to the Color Glass Condensate, but it also decays into quarks and gluons which are closer
in description to that of a Quark Gluon Plasma.  Hence its name.  This classical ensemble of flux tubes
can decay classically because
\begin{eqnarray}
               D_0 \vec{E} & = & \vec{D} \times \vec{B} \nonumber \\
               D_0 \vec{B} & = & \vec{D} \times \vec{E}  
\end{eqnarray}

Nevertheless, the decay of such flux tubes should generate observable effects in two particle correlations.\cite{fb}-\cite{bland}
Such effects appear to have been observed in the two particle correlations measured at RHIC,
and are the subject of my plenary session talk at this meeting.
\begin{figure}[htbp]
\begin{center}
\includegraphics[width=.50\textwidth] {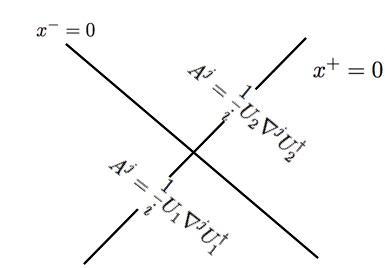} 
\end{center}
\caption{The vector potential corresponding to a single sheet of colored glass.}
\label{cgcpotential}
\end{figure}

How are these Glasma fields formed?  The initial vector potential corresponding to single sheet of colored glass is shown in Fig. \ref{cgcpotential}.  If we choose a field which is a pure two dimensional
gauge transformation of vacuum, but a different gauge transform on either side of the sheet located at $x^- = 0$, the Yang-Mills equations are solved for $x^- \ne 0$.  There is a discontinuity at $x^- = 0$,
and this discontinuity in the Yang-Mills equations generates the sources of charges corresponding to
the Lorentz contracted hadron.  The color electric and color magnetic fields generated by this
vector potential exist only on the sheets and have $\vec{E} \perp \vec{B} \perp \hat{z}$.

\begin{figure}[htbp]
\begin{center}
\includegraphics[width=.60\textwidth] {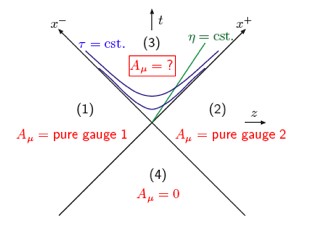} 
\end{center}
\caption{The vector potential corresponding to the collisions of two sheets of colored glass.}
\label{twopotential}
\end{figure}

Now consider the collision of two sheets of colored glass as shown in Fig. \ref{twopotential}.
In the backward lightcone, we use overall gauge freedom to specify $A^\mu = 0$.  Within the side light cones,
we have $A_1$ and $A_2$ chosen so that the correct sources for the hadron are generated along the backward light cone.  In the forward light cone we can have the Yang-Mills equation properly generate the source if infinitesimally near the light cone, we take $A = A_1 + A_2$.  However, since the sum of two gauge  fields which are gauge transforms of vacuum are not a a gauge transform of vacuum for non-abelian theories, this field must be taken as an initial condition, so that the fields evolve classically into the forward light cone.  This is how the Glasma fields are made, which eventually evolve into the Quark Gluon Plasma.

The fields generated in this way are very slowly varying in rapidity.  All of the variation ultimately arises because of the renormalization group evolution of the sources.  Therefore, the longitudinal Glasma fields are long range in rapidity.  The origin of these long range fields is seen from the Yang-Mills equations,
which near the forward light cone become. \begin{eqnarray}
        \nabla \cdot E_{1,2} & = & A_{2,1} \cdot E_{1,2} \nonumber \\
        \nabla \cdot B_{1,2} & = & A_{2,1} \cdot B_{1,2}
\end{eqnarray}
It can be shown that the color electric charge and color magnetic charge densities on each sheet are equal in magnitude but opposite in sign.  On the average,
there is as much strength in color electric field as there is in color magnetic because of the
symmetry of the Yang-Mill equation under $E \leftrightarrow B$, and the symmetry of the fields in the Color Glass Condensate.

The fields associated with the Glasma will develop turbulent or chaotic modes.\cite{mrocz}-\cite{fukushima}
This means that
if one starts with the boost invariant field described above, then in a time scale of order
$t \sim ln(1/\alpha_S)/Q_{sat}$ small fluctuations which are not boost invariant will grow and begin to dominate the system,  This becomes a large effect after the bulk of the Glasma have evaporated,
$t \sim 1/Q_{sat}$.  there should nevertheless remain a significant component of classical fields
until a time of order $ t \sim 1/(\alpha_sQ_{sat})$.  These instabilities might or might not have something to do with the early thermalization seen at RHIC.  They might also generate contributions to transverse flow at early times.  This problem is an amusing one since these early time fluctuations may ultimately
generate a Kolmogorov spectrum of fluctuations. 

\subsection{Total Multiplicity in Au-Au Collisions at RHIC}
 
 \begin{figure}[htbp]
\begin{center}
\includegraphics[width=.60\textwidth] {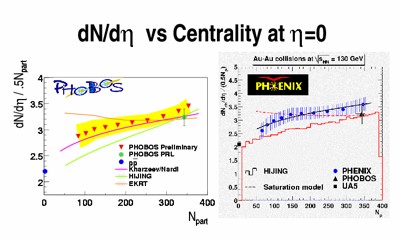} 
\end{center}
\caption{The multiplicity of produced particles in heavy ion collisions compared to CGC-Glasma expectations}
\label{rhicmultiplicity}
\end{figure}

One of the early successes of the CGC-Glasma description was the computation of the rapidity density of produced particles in RHIC Au-Au collisions.\cite{venugopalan},\cite{hirano}  The CGC-Glasma provided one of the very few correct predictions for the multiplicity, and as shown in Fig.  \ref{rhicmultiplicity}, correctly predicted the dependence of multiplicity upon centrality (and energy). 

\subsection{Event by Event P and CP Violation}

It has been suggested that in RHIC collisions, there might be large event by event CP and P violations.
\cite{kharzeevcp}-\cite{harmen}
This can occur by topological charge changing processes generated in the time evolution of the Glasma.
These are analogous to the sphaleron transitions which generate electroweak-baryon number violation in the standard model at temperatures of the order of the electroweak scale.\cite{manton}-\cite{arnoldmclerran}  In QCD, they are associated with anomalous helicity flip processes.  Such processes an occur at very early times during
the Glasma phase of evolution.

\begin{figure}[htbp]
\begin{center}
\includegraphics[width=.40\textwidth] {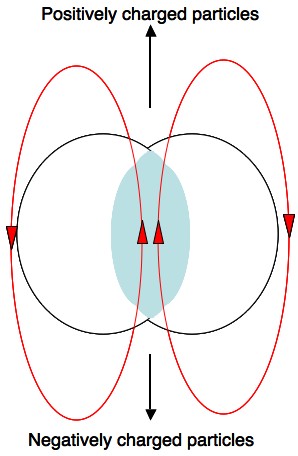} 
\end{center}
\caption{The chiral magnetic effect as induced by the off impact parameter zero collisio
of two heavy ions.}
\label{chiral}
\end{figure}

Of course such helicity flip processes take place all the time in the QCD vacuum.
since instanton processes are common-place when the coupling is of order one.  What makes heavy ion collisions special is that early in the collision, the electromagnetic charges of the nuclei can generate a strong electromagnetic field
in off impact parameter zero collisions, and that the high energy density of the Glasma fields makes the coupling small so that effects are computable.  The magnetic field decays rapidly in a time of order $1/Q_{sat}$.  If there is a net helicity induced by topological charge changing
processes in the presence of a magnetic field, then one generates an electromagnetic current parallel to the magnetic field.  This is because the magnetic moments align in the magnetic field, and since there is net helicity, this in turn implies a net current.  In this way, by measuring the fluctuations in the current, ome might measure underlying topological charge changing processes.  This effect is called the Chiral Magnetic effect, and is illustrated in Fig. \ref{chiral}

\section{Summary and Conclusions}

The Color Glass Condensate and the Glasma are forms of matter predicted by QCD.  They provide a successful phenomenology of high energy hadron-hadron and lepton-hadron-collisions.  At RHIC and Hera energies, they provide a semi-quantitative framework which describes a wide variety
of different processes.  The theory underlying the CGC and the Glasma  becomes best at the highest energy where the saturation momentum is large and the interaction strength of QCD at that scale $\alpha_s(Q_{sat})$ is correspondingly small.  Clearly, experiments at LHC energies, can test these ideas.  Another way to make the saturation momentum large is of course to use lepton-nucleus collisions where the density is enhanced because of the nucleus.\cite{erhic}

\section{Acknowledgments}  I thank the organizers of the International Symposium on Miltiparticle Dynamics for their kind hospitality.  

This manuscript has been authorized under Contract No. DE-AC02-98CH0886 with the US Department of Energy.

\end{document}